\documentclass[twocolumn,trackchanges]{aastex631}

\begin{document}

\title{On the Contribution of Unresolved Pulsars to the Ultra-high-energy Galactic Diffuse Gamma-Ray Emission}

\author{Samy Kaci}
\correspondingauthor{Samy Kaci}
\email{samykaci@sjtu.edu.cn}
\affiliation{Tsung-Dao Lee Institute, Shanghai Jiao Tong University\\ Shanghai 201210, P. R. China}
\affiliation{School of Physics and Astronomy, Shanghai Jiao Tong University\\ Shanghai 200240, P. R. China}


\author{Gwenael Giacinti}
\correspondingauthor{Gwenael Giacinti}
\email{gwenael.giacinti@sjtu.edu.cn}
\affiliation{Tsung-Dao Lee Institute, Shanghai Jiao Tong University\\ Shanghai 201210, P. R. China}
\affiliation{School of Physics and Astronomy, Shanghai Jiao Tong University\\ Shanghai 200240, P. R. China}

\author{Dmitri Semikoz}
\affiliation{APC, Université Paris Cité, CNRS/IN2P3, CEA/IRFU, Observatoire de Paris\\ 119 75205 Paris, France}



\begin{abstract}

The ultra high-energy (UHE) Galactic diffuse gamma-ray emission holds important information on the propagation of cosmic rays in the Galaxy. However, its measurements suffer from a contamination from unresolved sources whose contribution remains unclear. In this Letter, we propose a novel data-driven estimate of the contribution of unresolved pulsar wind nebulae and TeV halos based on the information present in the ATNF and the LHAASO catalogs. We find that in the inner Galaxy, this contribution is limited to $\sim38\%\pm10\%$ of the diffuse flux measured by LHAASO at $\sim20\,\rm{TeV}$ in the case where all sources associated to pulsars contribute as unresolved sources, and this fraction drops with energy to less than $21\%\pm6\%$ above $100\,\rm{TeV}$. In the outer Galaxy, this contribution is always subdominant. In particular, it reaches at most $\sim18\%\pm2\%$ at $10\,\rm{TeV}$ and is less than $\sim7\%\pm1\%$ above $\sim25\,\rm{TeV}$. We conclude that the UHE Galactic diffuse gamma-ray emission cannot be dominated by unresolved pulsar sources above a few tens of $\rm{TeV}$.

\end{abstract}

\keywords{astroparticle physics --- (stars:) pulsars: general --- gamma rays: diffuse background}


\section{Introduction}\label{sec:intro}
The UHE Galactic diffuse gamma-ray background is the gamma-ray emission resulting from the interactions of the sea of $\rm{TeV}$--$\rm{PeV}$ cosmic rays with different components of the interstellar medium. Its study can help to constrain the properties of cosmic-ray propagation in the interstellar medium, and of the Galactic magnetic field. It can even serve as a proxy to constrain the number and the nature of the still unknown sources of $\rm{PeV}$ cosmic rays in our Galaxy (the so-called PeVatrons), as we have shown in \cite{me}.

In the $\sim 10\,\rm{TeV}-\rm{PeV}$ range, the  leptonic contribution to the true Galactic diffuse gamma-ray background is expected to be smaller than $5$\% \citep{lipari}.
Because of the short cooling time of PeV electrons via synchrotron radiation ($\sim0.6\,$kyr), they remain confined within $\sim200\,$pc around their sources and are unable to generate an UHE sea of leptonic cosmic rays. As a result, they can only contribute locally to the UHE Galactic gamma-ray emission as resolved/unresolved leptonic sources, while its truly diffuse component is from a hadronic origin. Nevertheless, measurements of the diffuse gamma-ray background still suffer from a contamination from unresolved leptonic sources whose contribution remains poorly constrained. In particular, estimates of the contamination from unresolved pulsars strongly depend on the assumptions used to derive it, and range from small \citep[e.g.][]{kefeng} to dominant in  \cite{neutrinos_tevhalo} where a contribution from TeV halos reaching $\sim60\%$ at $10\,$TeV has been reported, while \cite{Martin2022} found that it depends on the unknown fraction of middle-aged (from $\sim30$ to $\sim400\,$kyr) pulsars that are able to develop a TeV halo. Moreover, the Large High Altitude Air Shower Observatory (LHAASO) collaboration reported its measurement of the diffuse gamma-ray flux from $10\,\rm{TeV}$ to $1\,\rm{PeV}$ in \cite{lhaaso_diffuse} where they claim to have found a flux $2\sim3$ times higher than their expectations, while AS$\gamma$, another gamma-ray observatory, reported a diffuse flux in \cite{as_gamma_diffuse} substantially higher than that of LHAASO. As a result, the knowledge of a robust estimate of the contribution of unresolved sources has become crucial in order to interpret observational data.

In this Letter, we aim to provide such an estimate for the population of unresolved pulsars. More specifically, we investigate the contribution of unresolved pulsar wind nebulae (PWNe) and TeV halos, which have already been suggested as potential candidates contributing to the diffuse flux of LHAASO \citep{lhaaso_diffuse} and of the Fermi Large Area Telescope (Fermi-LAT) at $\rm{GeV}$ energies \citep{vittoria}. In what follows, we shall not distinguish between pulsars, PWNe and TeV halos for convenience.

We adopt a data-driven approach based on the Australia Telescope National Facility (ATNF) and LHAASO catalogs. While we find that in the region limited by $125^{\circ}\leq l<235^{\circ}$ and $|b|\leq5^{\circ}$ , referred to as the outer Galaxy in \cite{lhaaso_diffuse}, the contribution of unresolved pulsars is always smaller than $18\%\pm2\%$, we observe that in the inner Galaxy ($15^{\circ}\leq l<125^{\circ}$ and $|b|\leq5^{\circ}$), it depends on our assumption on the minimum pulsar spindown luminosity required to power an UHE gamma-ray emission. Nevertheless, this contribution is limited to $\sim38\%\pm10\%$ of the flux reported by LHAASO around $20\,\rm{TeV}$ and decreases with energy.

This Letter is organized as follows. In section \ref{sec:model} we present our model for the generation of the synthetic population of pulsars. In section \ref{sec:results} we present and interpret our results. Finally, we confront our results to those present in the literature in section \ref{sec:discussion} and share our conclusions in section \ref{sec:conclusion}.

\section{Methods}\label{sec:model}
\begin{figure*}[t!]
    \centering
    \includegraphics[scale=0.85]{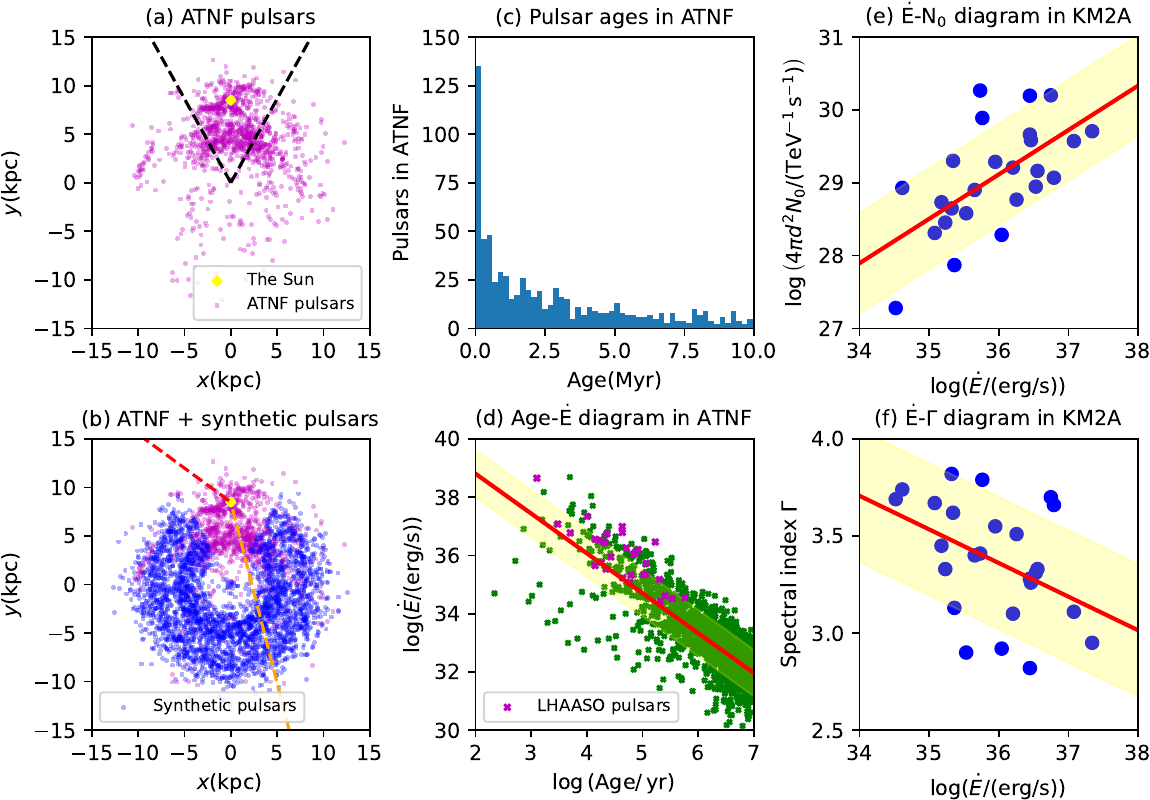}
    \caption{Summary of the source generation procedure. Panel (a) shows the ATNF pulsars with age from $0$ to $10\,\rm{Myr}$ in the region (between the two dashed lines) used to extract their statistical properties. Panel (b) shows the ATNF and the synthetic pulsars generated with the same statistical properties and the field of view of LHAASO starting from the orange dashed line and ending at the red dashed line (counterclockwise). Histogram (c) shows the age distribution of the ATNF pulsars in the sampling region shown in panel (a). Panel (d) shows the relationship between the age of pulsars and their spindown power. Panels (e) and (f) respectively show the relationship between the spindown power and the reference flux and the spindown power and the spectral index reported by KM2A in \cite{lhaaso_catalog}. The red lines and the yellow shaded areas in panels (d), (e) and (f) respectively show the linear regression fit to the data and the 68\% containment region of the data or the $1\sigma$ deviation from the fit.}
    \label{fig:relationships}
\end{figure*}

We aim to provide a robust estimate of the contribution of pulsars to the UHE Galactic diffuse gamma-ray emission. To do so, we choose to rely on data and avoid invoking any theoretical assumption whenever possible. In order to generate our population of pulsars and their gamma-ray emission, we use the statistical information available in the ATNF and LHAASO catalogs as described in figure \ref{fig:relationships}.

We first select pulsars with age ranging from $0$ to $10\,\rm{Myr}$ from the ATNF catalog {\tt\string version 2.1.1}\footnote{\url{https://www.atnf.csiro.au/research/pulsar/psrcat/index.html?version=2.1.1}} \citep{atnf_1}. Considering the contribution from pulsars older than $10\,\rm{Myr}$ does not strongly affect our results below because of their small number and spindown power. In order to derive a spatial distribution of pulsars, assumed to be axi-symmetric, we only consider pulsars located in the cone starting at the Galactic center and limited by the two dashed lines forming an angle of $60^{\circ}$ shown in panel (a) of figure \ref{fig:relationships}. We consider that the ATNF catalog is sufficiently complete in this region to be representative of the whole Galaxy. After that, we divide this region into $13$ portions of rings (bins) of $1\,\rm{kpc}$ of thickness, and count the number of pulsars detected in each bin. We exclude pulsars with unknown distance or altitude $z>1.7\,\rm{kpc}$ in order to avoid the inclusion of extra-Galactic pulsars and those which are outside the LHAASO Galactic plane survey limited by $|b|\leq5^{\circ}$. Once the number of pulsars in each bin is determined, we assume that the rest of the Galaxy has the same structure as our sampling region and put additional pulsars randomly generated on the top of the ATNF pulsars until each part of the Galaxy is filled. More specifically, we fill each cone of $60^{\circ}$ opening angle and centered at the Galactic center with synthetic pulsars until the total number of pulsars in each of them reaches approximately $714$, with $714$ being the number of pulsars in the sampling region. We allow this number to fluctuate following a Gaussian distribution with a standard deviation $\sigma\simeq\sqrt{714}$. For the $z$-distribution of pulsars, we have tested a Gaussian centered around $0$ with a standard deviation of $70\,\rm{pc}$, which corresponds to the distribution of their progenitors, together with distributions broadening with pulsar age, taking into account their kick velocities. We tried different values up to 300\,km\,s$^{-1}$ for the average kick velocities of pulsars, and we did not observe any significant impact on our results. Therefore, in the following, we show the results for the former Gaussian distribution in $z$. The resulting population of pulsars is shown in the panel (b) of figure \ref{fig:relationships}. This gives a total of $\sim4300$ pulsars likely to contribute to the UHE Galactic gamma-ray emission.

We generate the age of pulsars following the age distribution of ATNF in our sampling region, as shown in the histogram (c) of figure \ref{fig:relationships}. We simply divide the ages of the sampling region into $10$ bins in age between $0$ and $10\,\rm{Myr}$, each having $\sim70$ pulsars. Then, we randomly assign ages to pulsars in the rest of the Galaxy requiring that, in each cone, all age bins contain approximately the same number of pulsars as in the sampling region, the differences being due to the fluctuations in the number of generated pulsars. We have checked that the age distribution displayed in histogram (c) of figure \ref{fig:relationships} is independent from the distance to the Earth and is representative of the whole Galaxy.

In order to generate the spindown power of the synthetic pulsars we use the fitting expression:
\begin{equation}\label{edot}
    \log\left(\frac{\dot{E}}{\rm{erg}/\rm{s}}\right) = 41.54 - 1.37\log\left(\frac{\tau}{\rm{yr}}\right)
\end{equation}
where $\dot{E}$ represents the spindown power of pulsars and $\tau$ their age. Eq. (\ref{edot}) is obtained through a linear regression on the data present in ATNF for the entire Galaxy. In log-scale the spindown power of a pulsar is roughly inversely proportional to its age. However, it also depends on other parameters such as the initial period of the pulsar. Such additional parameters induce an important (model-dependent) scattering of the data around the linear fitting function, as shown in panel (d) of figure \ref{fig:relationships}. We account for this scattering by adding a Gaussian noise whose intensity is tuned to match the correlation coefficient of the synthetic data with that of the ATNF data. This results in a standard deviation $\sigma = 0.8$ for the noise.

Finally, we generate a gamma-ray emission for each pulsar with a common physical extension assumed to be $20\,\rm{pc}$  \citep[to match the extension of Geminga and Monogem,][]{gg} and a gamma-ray spectrum defined as in the LHAASO catalog \citep{lhaaso_catalog}. Thus, we assume that all spectra are given by:
\begin{equation}\label{sepctrum}
    dN/dE \equiv N_0\left(E/E_0\right)^{-\Gamma}
\end{equation}
where the flux $dN/dE$ is given in $\rm{TeV}^{-1}\,\rm{cm}^{-2}\,\rm{s}^{-1}$, $E_0 = 50\,\rm{TeV}$ is the reference energy, the reference flux $N_0$ is given in $\rm{TeV}^{-1}\,\rm{cm}^{-2}\,\rm{s}^{-1}$ and $\Gamma$ is the spectral index. To determine the reference fluxes of sources we use the relation:
\begin{equation}\label{ref_flux}
    \log\left(\frac{4\pi d^2N_0}{\rm{TeV}^{-1}\,\rm{s}^{-1}}\right)=7.19+0.61\log\left(\frac{\dot{E}}{\rm{erg}/\rm{s}}\right)
\end{equation}
where $d$ represents the distance of the source to the Earth. This relation gives the quantity $4\pi d^2N_0$ which represents the true luminosity of the source. As previously, it is obtained through a linear regression performed on the Square Kilometer Array (KM2A) data in the LHAASO catalog after removing the pulsars which do not have a distance and the Crab whose exceptional properties may bias the generation of the random gamma-ray emissions. These data are represented in the panel (e) of figure \ref{fig:relationships}. The reference flux that would be measured by KM2A can be retrieved by removing the factor $4\pi d^2$. In order to account for the uncertainties, here again we have added a Gaussian noise with a standard deviation $\sigma = 0.69$, which ensures that the correlation coefficient of the synthetic data matches that of the experimental data. A similar relation between the spectral index and the spindown power is obtained through the same procedure and is given by:
\begin{equation}\label{ref_index}
    \Gamma = 9.59-0.17\log\left(\frac{\dot{E}}{\rm{erg}/\rm{s}}\right)
\end{equation}
with the relevant data associated to the linear regression being represented in the panel (f) of figure \ref{fig:relationships}. Several other parameters, such as the geometry of the magnetic field or the multiplicity factor of electron-positron pairs, may play an important role in particle acceleration within the source. Their contribution is reflected in the scattering of the data visible in panels (e) and (f) of figure \ref{fig:relationships}. Here again we account for the scattering in panel (f) by adding a Gaussian noise with a standard deviation $\sigma = 0.34$ leading to a correlation coefficient for the synthetic data similar to that of the data of KM2A.

Following the previous procedure, we generate lists of sources in which the flux of each individual source is limited by the sensitivity of KM2A. When concluding on whether a source can be detected or not, we take into account the exposure of different regions of the sky by introducing a declination-dependent attenuation factor following \cite{attenuation}.

\begin{figure*}[t!]
    \centering
    \includegraphics[width=\textwidth]{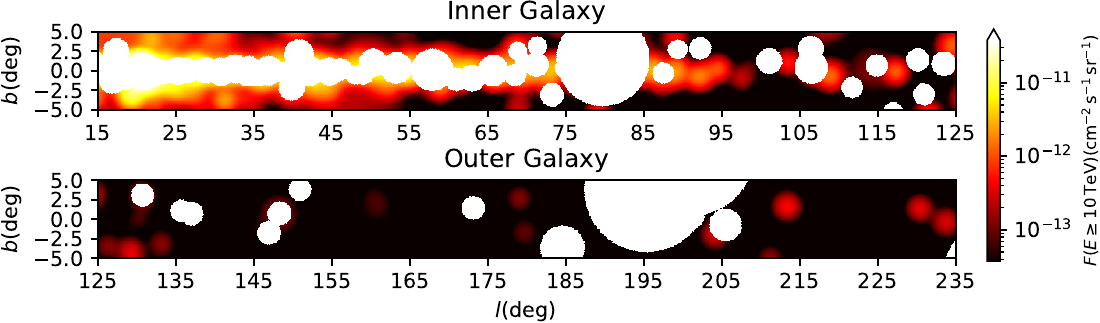}
    \caption{Expected gamma-ray flux of unresolved \textbf{pulsars} (without cut in spindown power) above $10\,\rm{TeV}$ in the $\left(l, b\right)$ plane for the inner Galaxy (upper panel) and outer Galaxy (lower panel). The white regions represent the masks of LHAASO from \cite{lhaaso_diffuse}. The upper bound of the color-bar represents the sensitivity of LHAASO to a point source located at a declination of $-10^{\circ}$ and the lower bound of the color-bar represents $0.1\%$ of its upper bound.}
    \label{fig:skymap}
\end{figure*}

In order to compute the diffuse flux, we use the same masking procedure as LHAASO in \cite{lhaaso_diffuse} and convolve our source extensions with the point-spread function (PSF) of KM2A at $10\,\rm{TeV}$ \citep{angular_resolution}. The gamma-ray flux resulting from this procedure is illustrated in figure \ref{fig:skymap}, together with the masks used by LHAASO. By construction, this represents the flux of unresolved pulsars above $10\,\rm{TeV}$ for the inner and the outer Galaxy. The upper bound of the color-bar represents the sensitivity of LHAASO to a point source (assuming a Gaussian profile) located at a declination of $-10^{\circ}$ and the lower bound of the color-bar represents $0.1\%$ of its upper bound.

To perform our analysis we consider three cases: (i) We assume that all pulsars can power an UHE gamma-ray emission without any restriction. This serves as a limiting case. (ii) We exclude pulsars with $\dot{E}\leq3\times10^{33}\,\rm{erg}\,\rm{s}^{-1}$ which corresponds to the smallest spindown power for gamma-ray emitting pulsars reported by Fermi-LAT at $\rm{GeV}$ energies in \cite{fermi_catalog}. In this case, we make the assumption that it is unlikely that pulsars that are too weak to be detected at $\rm{GeV}$ energy may have an UHE emission sufficiently significant to impact the value of the diffuse gamma-ray background. (iii) Finally, we make our most stringent cut and exclude pulsars with $\dot{E}\leq10^{34}\,\rm{erg}\,\rm{s}^{-1}$. This cut reflects the fact that the spindown of the weakest pulsar (in terms of spindown power) detected by LHAASO is $\sim2.7\times10^{34}\,\rm{erg}\,\rm{s}^{-1}$.

In order to evaluate the contribution of unresolved pulsars to the diffuse flux measured by LHAASO, we have run 1000 simulations. In each simulation, we generate a population of synthetic pulsars on top of the ATNF ones. We then randomly assign a spindown power and a gamma-ray spectrum to each source using equations (\ref{edot}), (\ref{ref_flux}) and (\ref{ref_index}) where the uncertainty in the fits is taken into account through the random noise added to the linear relationship. After that, we compute the resulting gamma-ray flux following the same procedure as LHAASO.

\section{Results}\label{sec:results}
\begin{figure*}[b!]
    \centering
    \includegraphics[width=\textwidth]{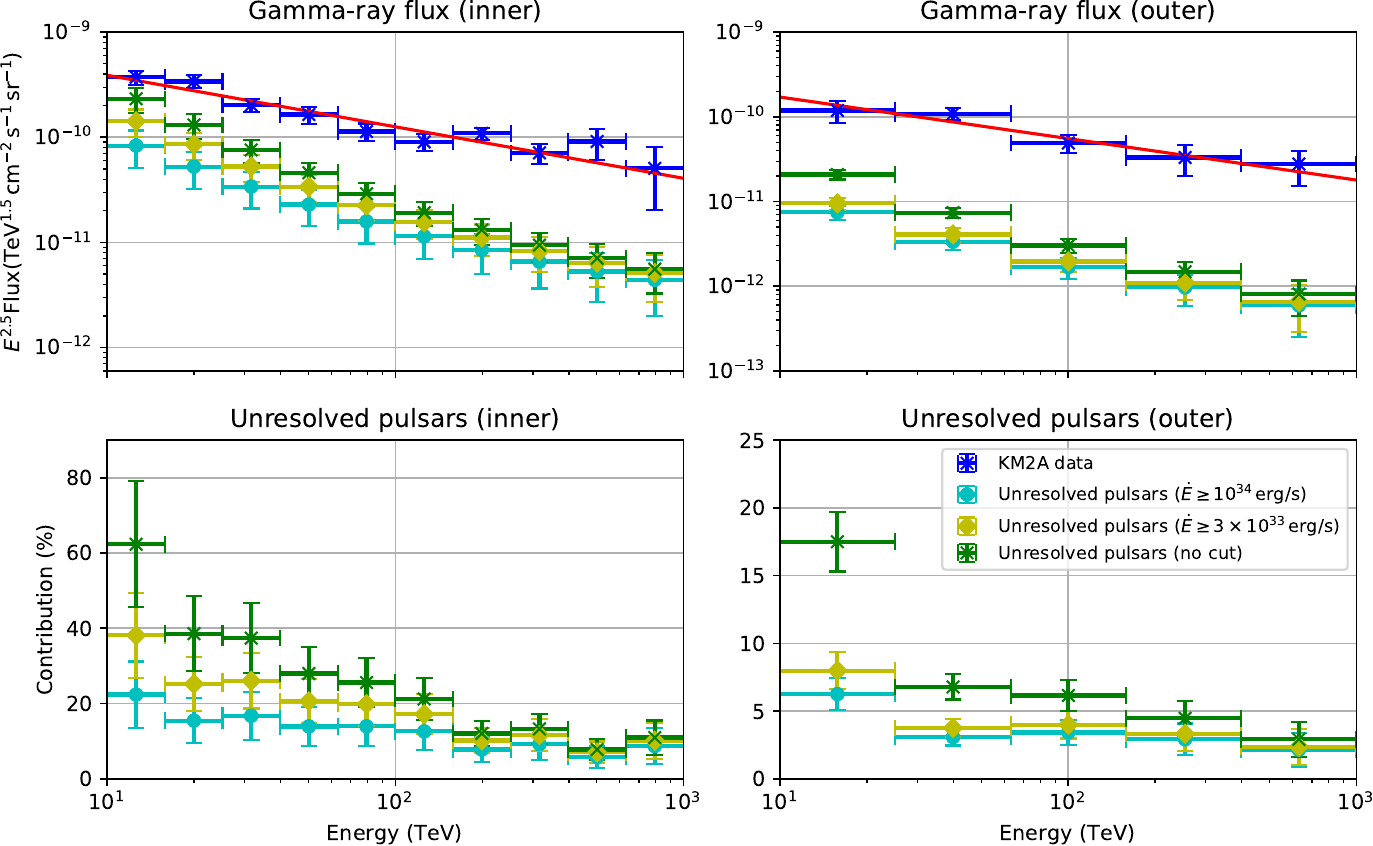}
    \caption{Contribution of unresolved pulsars to the diffuse gamma-ray background measured by LHAASO. The upper left (right) panel shows the diffuse flux measured by KM2A and the flux of unresolved pulsars in the inner (outer) Galaxy. The lower left (right) panel shows the relative contribution of unresolved pulsars to the diffuse flux measured by LHAASO in the inner (outer) Galaxy.}
    \label{fig:results1}
\end{figure*}

We present our results related to the contribution of unresolved pulsars to the diffuse background measured by LHAASO in figure \ref{fig:results1}.

The upper left (right) panel of figure \ref{fig:results1} shows the data points of KM2A in \cite{lhaaso_diffuse} together with the flux contributed by our synthetic population of pulsars in the inner (outer) Galaxy. The lower left (right) panel shows the relative contribution of the same population of pulsars to the flux measured by KM2A in the inner (outer) Galaxy. The data points relative to the unresolved pulsars represent the average over $1000$ simulations and their error bars represent one standard deviation. 

Concerning the outer Galaxy, the right panels of figure \ref{fig:results1} show that the contribution of unresolved pulsars is smaller than $\sim18\%\pm2\%$ at all energies and is even smaller than $\sim7\%\pm1\%$ above $\sim25\,$TeV. This result does not depend on a specific cut in the value of the spindown power and is due to the scarcity of sources in the outer Galaxy.

From the left panels of figure \ref{fig:results1} we see that above $100\,\rm{TeV}$ the contribution of unresolved pulsars to the diffuse flux is smaller than $\sim21\%\pm6\%$ of the flux of LHAASO and is, at most, weakly dependent on the value of the spindown power adopted for the cut. In contrast, in the part of the spectrum from $10$ to $100\,\rm{TeV}$, the level of contamination strongly depends on the value of the spindown power chosen for the cut and varies between $\sim22\%\pm9\%$ and $\sim63\%\pm17\%$ of the flux reported by LHAASO at $10\,\rm{TeV}$. In addition, it can be seen that in the context of the LHAASO data and interpretations, the cut at $\dot{E}\gtrsim10^{34}\,\rm{erg}\,\rm{s}^{-1}$ is disfavored, as it cannot account for their reported excess, while the two other scenarios remain viable. In all cases, it seems unlikely that more than $\sim38\%\pm10\%$ of the diffuse flux can be attributed to the population of unresolved pulsars above $\sim20\,$TeV. This could happen only in a small number of nonphysical realizations leading to an excessively large contribution, surpassing $100\%$ of the total flux in the most extreme cases. These realizations correspond to very rare configurations of the Galaxy.

\begin{figure*}
    \centering
    \includegraphics[width=\textwidth]{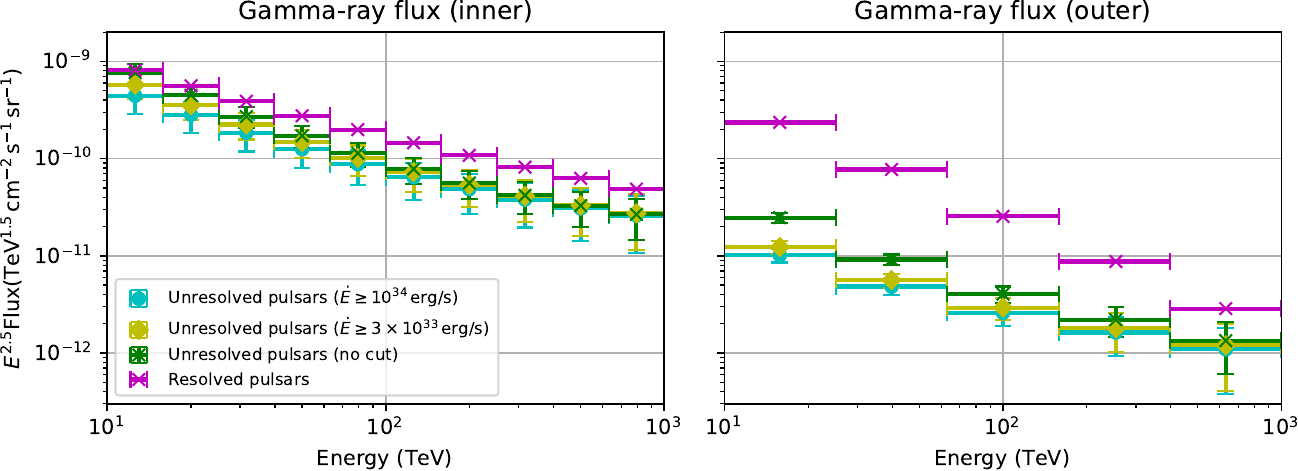}
    \caption{Flux of unresolved pulsars together with that of resolved pulsars without masks in the inner (left panel) and outer (right panel) Galaxy. The same cuts in spindown power as previously have been used.}
    \label{fig:results2}
\end{figure*}

An important quantity constrained in this Letter is the flux of unresolved pulsars in the entire Galactic plane. While the masking procedure was introduced to avoid contamination from the population of resolved sources (both leptonic and hadronic), it also removes the contribution of unresolved sources that are in the masked regions. However, if the masks are removed to include the information from the Galactic plane, it is expected that the contaminating flux of unresolved sources will grow significantly. In figure \ref{fig:results2}, we provide an estimate of the flux of unresolved pulsars without masks.

The left (right) panel of figure \ref{fig:results2} shows the flux of unresolved pulsars together with that of resolved pulsars in the inner (outer) Galaxy in the case where no masking procedure is adopted. Here again, the Crab is excluded. Both panels show that in the case where the information from the whole Galactic plane is taken into account, the flux of unresolved pulsars is higher. Moreover, the left panel shows that in the inner Galaxy, the slope of the flux of unresolved pulsars is very close to that of resolved pulsars, which confirms \textit{a posteriori} the self-consistency of our approach. Concerning the outer Galaxy, the slope of unresolved pulsars is also very close to that of the resolved pulsars of the inner Galaxy. However, it is harder than that of the resolved pulsars of the outer Galaxy. This is due to the fact that very few sources have been resolved there, leading to higher statistical fluctuations in the region. The fact that the spectrum of unresolved pulsars is softer than that of the diffuse background in figure \ref{fig:results1} is anyway compatible with our finding that the contamination from unresolved pulsars decreases with energy.

\section{Discussion}\label{sec:discussion}
In this Letter, we have investigated the contribution of unresolved pulsars to the Galactic diffuse gamma-ray flux. Our approach leads to a contribution in the inner Galaxy ranging from $\sim22\%\pm9\%$ to $\sim63\%\pm17\%$ of the flux of LHAASO around $10\,\rm{TeV}$, less than $\sim38\%\pm10\%$ above $\sim20\,\rm{TeV}$, and less than $\sim21\%\pm6\%$ above $100\,\rm{TeV}$. It is always below $\sim18\%\pm2\%$ in the outer Galaxy. 

Our results would still hold if a high-energy cutoff were to be introduced in eq. (\ref{sepctrum}), as this would only decrease the already very small high-energy contribution of unresolved pulsars. In addition, we find that the approach described in section \ref{sec:model} leads to a spectrum for unresolved pulsars similar to that of resolved pulsars, confirming the self-consistency of our work.

Since our study is data-driven, we follow the actual, non-uniform age distribution of ATNF pulsars shown in figure \ref{fig:relationships}. Another possibility considered by other authors, is to assume a uniform age distribution, with e.g. a birthrate of $0.7$ pulsars per century as in \cite{tevhalos_let}. However, doing so leads to model-dependent results substantially different from ours. Albeit not data-driven, we have also investigated this scenario by adding an artificial population of old pulsars to the data consistent with a constant Galactic pulsar birthrate of $\sim5\,\rm{kyr}^{-1}$. We find that this would violate LHAASO observations. In particular, there would be $\sim350-1300$ pulsars detectable by LHAASO in our simulations, which is clearly at odds with the results reported in the LHAASO catalog \citep{lhaaso_catalog}. In addition, we find that the vast majority of pulsars observed by Fermi-LAT at $\rm{GeV}$ energies in the vicinity of Earth (distance $\lesssim4\,\rm{kpc}$) have a radio counterpart in ATNF and display the same age distribution as in ATNF for the same region, which further strengthens our choice to follow the ATNF distribution in age.

Our results are also in tension with those of \cite{neutrinos_tevhalo} except for the high energy part of the outer Galaxy. In particular, the authors find that, in the inner Galaxy, unresolved pulsars account for more than 50\% of the LHAASO flux in most of its energy range. Moreover, the fact that this contribution almost does not decrease with energy, suggests that its sepectrum is as hard as that of the diffuse background, which is in conflict with the steeper spectrum of the resolved LHAASO pulsars. These tensions are due to the very different approaches adopted. We mostly rely on data avoiding as much as possible the inclusion of model-dependent theoretical hypotheses, while \cite{neutrinos_tevhalo} have adopted a more theoretical approach, such as assuming a unique injection index $s = 2.2$ for electron/positron pairs.

Finally, we note that our results are compatible with the Galactic neutrino flux of IceCube \citep{icecube}. Indeed, our $\sim38\%\pm10\%$ upper limit on the pulsar contribution to the diffuse flux above $20\,$TeV calls for a sizable additional contribution to it, leaving ample room for a (large) hadronic component.

\section{Conclusion}\label{sec:conclusion}
In conclusion, we have provided a new estimate of the contribution of unresolved pulsars, which are the most prominent UHE leptonic sources, to the Galactic diffuse gamma-ray background. Unlike previous studies which relied on theoretical assumptions, we adopted here a self-consistent data-driven approach. We notably provided stringent constraints on the contribution of unresolved pulsars at different energies in different parts of the sky. In particular, we find the pulsar contribution to be smaller than $\sim38\%\pm10\%$ at $20\,\rm{TeV}$ and smaller than $\sim21\%\pm6\%$ above $100\,\rm{TeV}$. While a contribution from other sources such as supernova remnants or micro-quasars cannot be ruled out, this information allows to infer robust upper limits on the hadronic component of the diffuse background, whose characterization will be crucial to constrain cosmic-ray propagation, and ultimately unveil the nature of the long-sought-after sources of $\rm{PeV}$ cosmic rays.

\begin{acknowledgments}
The authors acknowledge the use of the ATNF catalog. Samy Kaci acknowledges funding from the Chinese Scholarship Council (CSC). This work is supported by the National Natural Science Foundation of China under Grants Nos. 12350610239, and 12393853.
\end{acknowledgments}

\bibliography{sample631}{}

\begin{thebibliography}{}
\expandafter\ifx\csname natexlab\endcsname\relax\def\natexlab#1{#1}\fi
\providecommand{\url}[1]{\href{#1}{#1}}
\providecommand{\dodoi}[1]{doi:~\href{http://doi.org/#1}{\nolinkurl{#1}}}
\providecommand{\doeprint}[1]{\href{http://ascl.net/#1}{\nolinkurl{http://ascl.net/#1}}}
\providecommand{\doarXiv}[1]{\href{https://arxiv.org/abs/#1}{\nolinkurl{https://arxiv.org/abs/#1}}}

\bibitem[{Abbasi {et~al.}(2023)Abbasi, Ackermann, Adams, Aguilar, Ahlers, Ahrens, Alameddine, Alves, Amin, Andeen, Anderson, Anton, Argüelles, Ashida, Athanasiadou, Axani, Bai, V., Barwick, Basu, Baur, Bay, Beatty, Becker, Tjus, Beise, Bellenghi, Benda, BenZvi, Berley, Bernardini, Besson, Binder, Bindig, Blaufuss, Blot, Boddenberg, Bontempo, Book, Borowka, Böser, Botner, Böttcher, Bourbeau, Bradascio, Braun, Brinson, Bron, Brostean-Kaiser, Burley, Busse, Campana, Carnie-Bronca, Chen, Chen, Chirkin, Choi, Clark, Clark, Classen, Coleman, Collin, Connolly, Conrad, Coppin, Correa, Cowen, Cross, Dappen, Dave, Clercq, DeLaunay, López, Dembinski, Deoskar, Desai, Desiati, de~Vries, de~Wasseige, DeYoung, Diaz, Díaz-Vélez, Dittmer, Dujmovic, Dunkman, DuVernois, Ehrhardt, Eller, Engel, Erpenbeck, Evans, Evenson, Fan, Fazely, Fedynitch, Feigl, Fiedlschuster, Fienberg, Finley, Fischer, Fox, Franckowiak, Friedman, Fritz, Fürst, Gaisser, Gallagher, Ganster, Garcia, Garrappa, Gerhardt, Ghadimi, Glaser, Glauch,
  Glüsenkamp, Goehlke, Goldschmidt, Gonzalez, Goswami, Grant, Grégoire, Griswold, Günther, Gutjahr, Haack, Hallgren, Halliday, Halve, Halzen, Minh, Hanson, Hardin, Harnisch, Haungs, Helbing, Henningsen, Hettinger, Hickford, Hignight, Hill, Hill, Hoffman, Hoshina, Hou, Huang, Huber, Huber, Hultqvist, Hünnefeld, Hussain, Hymon, In, Iovine, Ishihara, Jansson, Japaridze, Jeong, Jin, Jones, Kang, Kang, Kang, Kappes, Kappesser, Kardum, Karg, Karl, Karle, Katz, Kauer, Kellermann, Kelley, Kheirandish, Kin, Kiryluk, Klein, Kochocki, Koirala, Kolanoski, Kontrimas, Köpke, Kopper, Kopper, Koskinen, Koundal, Kovacevich, Kowalski, Kozynets, Krupczak, Kun, Kurahashi, Lad, Gualda, Lanfranchi, Larson, Lauber, Lazar, Lee, Leonard, Leszczyńska, Li, Lincetto, Liu, Liubarska, Lohfink, Mariscal, Lu, Lucarelli, Ludwig, Luszczak, Lyu, Ma, Madsen, Mahn, Makino, Mancina, Mariş, Martinez-Soler, Maruyama, McHale, McElroy, McNally, Mead, Meagher, Mechbal, Medina, Meier, Meighen-Berger, Merckx, Micallef, Mockler, Montaruli, Moore,
  Morik, Morse, Moulai, Mukherjee, Naab, Nagai, Nahnhauer, Naumann, Necker, Nguyen, Niederhausen, Nisa, Nowicki, Nygren, Pollmann, Oehler, Oeyen, Olivas, O'Sullivan, Pandya, Pankova, Park, Parker, Paudel, Paul, de~los Heros, Peters, Peterson, Philippen, Pieper, Pizzuto, Plum, Popovych, Porcelli, Rodriguez, Pries, Przybylski, Raab, Rack-Helleis, Raissi, Rameez, Rawlins, Rea, Rechav, Rehman, Reichherzer, Reimann, Renzi, Resconi, Reusch, Rhode, Richman, Riedel, Roberts, Robertson, Roellinghoff, Rongen, Rott, Ruhe, Ryckbosch, Cantu, Safa, Saffer, Salazar-Gallegos, Sampathkumar, Herrera, Sandrock, Santander, Sarkar, Sarkar, Satalecka, Schaufel, Schieler, Schindler, Schmidt, Schneider, Schneider, Schröder, Schumacher, Schwefer, Sclafani, Seckel, Seunarine, Sharma, Shefali, Shimizu, Silva, Skrzypek, Smithers, Snihur, Soedingrekso, Sogaard, Soldin, Spannfellner, Spiczak, Spiering, Stamatikos, Stanev, Stein, Stettner, Stezelberger, Stokstad, Stürwald, Stuttard, Sullivan, Taboada, Ter-Antonyan, Thwaites, Tilav,
  Tischbein, Tollefson, Tönnis, Toscano, Tosi, Trettin, Tselengidou, Tung, Turcati, Turcotte, Turley, Twagirayezu, Ty, Elorrieta, Valtonen-Mattila, Vandenbroucke, van Eijndhoven, Vannerom, van Santen, Veitch-Michaelis, Verpoest, Walck, Wang, Watson, Weaver, Weigel, Weindl, Weiss, Weldert, Wendt, Werthebach, Weyrauch, Whitehorn, Wiebusch, Willey, Williams, Wolf, Wrede, Wulff, Xu, Yanez, Yildizci, Yoshida, Yu, Yuan, Zhang, \& Zhelnin}]{icecube}
Abbasi, R., Ackermann, M., Adams, J., {et~al.} 2023, Science, 380, 1338, \dodoi{10.1126/science.adc9818}

\bibitem[{Amenomori {et~al.}(2021)Amenomori, Bao, Bi, Chen, Chen, Chen, Chen, Chen, Cirennima, Cui, Danzengluobu, Ding, Fang, Fang, Feng, Feng, Feng, Gao, Gou, Guo, Guo, He, He, Hibino, Hotta, Hu, Hu, Huang, Jia, Jiang, Jin, Kasahara, Katayose, Kato, Kato, Kawata, Kihara, Ko, Kozai, Labaciren, Le, Li, Li, Li, Lin, Liu, Liu, Liu, Liu, Liu, Lou, Lu, Meng, Munakata, Nakada, Nakamura, Nanjo, Nishizawa, Ohnishi, Ohura, Ozawa, Qian, Qu, Saito, Sakata, Sako, Shao, Shibata, Shiomi, Sugimoto, Takano, Takita, Tan, Tateyama, Torii, Tsuchiya, Udo, Wang, Wu, Xue, Yamamoto, Yang, Yokoe, Yuan, Zhai, Zhang, Zhang, Zhang, Zhang, Zhang, Zhang, Zhang, Zhao, Zhaxisangzhu, \& Zhou}]{as_gamma_diffuse}
Amenomori, M., Bao, Y.~W., Bi, X.~J., {et~al.} 2021, Phys. Rev. Lett., 126, 141101, \dodoi{10.1103/PhysRevLett.126.141101}

\bibitem[{Cao {et~al.}(2023)Cao, Aharonian, An, Axikegu, Bai, Bao, Bastieri, Bi, Bi, Cai, Cao, Cao, Cao, Chang, Chang, Chen, Chen, Chen, Chen, Chen, Chen, Chen, Chen, Chen, Chen, Chen, Chen, Cheng, Cheng, Cui, Cui, Cui, Cui, Dai, Dai, Dai, Danzengluobu, della Volpe, Dong, Duan, Fan, Fan, Fang, Fang, Feng, Feng, Feng, Feng, Feng, Gabici, Gao, Gao, Gao, Gao, Gao, Gao, Ge, Geng, Giacinti, Gong, Gou, Gu, Guo, Guo, Guo, Guo, Han, He, He, He, He, He, Heller, Hor, Hou, Hou, Hou, Hu, Hu, Hu, Huang, Huang, Huang, Huang, Huang, Huang, Huang, Ji, Jia, Jia, Jiang, Jiang, Jiang, Jin, Kang, Ke, Kuleshov, Kurinov, Li, Li, Li, Li, Li, Li, Li, Li, Li, Li, Li, Li, Li, Li, Li, Li, Li, Li, Li, Liang, Liang, Lin, Liu, Liu, Liu, Liu, Liu, Liu, Liu, Liu, Liu, Liu, Liu, Liu, Liu, Liu, Lu, Luo, Lv, Ma, Ma, Ma, Mao, Min, Mitthumsiri, Mu, Nan, Neronov, Ou, Pang, Pattarakijwanich, Pei, Qi, Qi, Qiao, Qin, Ruffolo, S\'aiz, Semikoz, Shao, Shao, Shchegolev, Sheng, Shu, Song, Stenkin, Stepanov, Su, Sun, Sun, Sun, Tam, Tang, Tang, Tian, Wang,
  Wang, Wang, Wang, Wang, Wang, Wang, Wang, Wang, Wang, Wang, Wang, Wang, Wang, Wang, Wang, Wang, Wang, Wang, Wang, Wang, Wei, Wei, Wei, Wen, Wu, Wu, Wu, Wu, Wu, Xi, Xia, Xia, Xiang, Xiao, Xiao, Xin, Xin, Xing, Xiong, Xu, Xu, Xu, Xu, Xue, Yan, Yan, Yan, Yang, Yang, Yang, Yang, Yang, Yang, Yang, Yang, Yang, Yao, Yao, Ye, Yin, Yin, You, You, Yu, Yuan, Yue, Zeng, Zeng, Zeng, Zha, Zhang, Zhang, Zhang, Zhang, Zhang, Zhang, Zhang, Zhang, Zhang, Zhang, Zhang, Zhang, Zhang, Zhang, Zhang, Zhang, Zhang, Zhang, Zhao, Zhao, Zhao, Zhao, Zhao, Zheng, Zhou, Zhou, Zhou, Zhou, Zhou, Zhou, Zhou, Zhu, Zhu, Zhu, Zhu, \& Zuo}]{lhaaso_diffuse}
Cao, Z., Aharonian, F., An, Q., {et~al.} 2023, Phys. Rev. Lett., 131, 151001, \dodoi{10.1103/PhysRevLett.131.151001}

\bibitem[{Cao {et~al.}(2024)Cao, Aharonian, An, Axikegu, Bai, Bao, Bastieri, Bi, Bi, Cai, Cao, Cao, Cao, Chang, Chang, Chen, Chen, Chen, Chen, Chen, Chen, Chen, Chen, Chen, Chen, Chen, Chen, Cheng, Cheng, Cui, Cui, Cui, Cui, Dai, Dai, Dai, Danzengluobu, della Volpe, Dong, Duan, Fan, Fan, Fang, Fang, Feng, Feng, Feng, Feng, Feng, Gabici, Gao, Gao, Gao, Gao, Gao, Gao, Ge, Geng, Giacinti, Gong, Gou, Gu, Guo, Guo, Guo, Guo, Han, He, He, He, He, He, Heller, Hor, Hou, Hou, Hou, Hu, Hu, Hu, Huang, Huang, Huang, Huang, Huang, Huang, Huang, Ji, Jia, Jia, Jiang, Jiang, Jiang, Jin, Kang, Ke, Kuleshov, Kurinov, Li, Li, Li, Li, Li, Li, Li, Li, Li, Li, Li, Li, Li, Li, Li, Li, Li, Li, Li, Liang, Liang, Lin, Liu, Liu, Liu, Liu, Liu, Liu, Liu, Liu, Liu, Liu, Liu, Liu, Liu, Liu, Lu, Luo, Lv, Ma, Ma, Ma, Mao, Min, Mitthumsiri, Mu, Nan, Neronov, Ou, Pang, Pattarakijwanich, Pei, Qi, Qi, Qiao, Qin, Ruffolo, Sáiz, Semikoz, Shao, Shao, Shchegolev, Sheng, Shu, Song, Stenkin, Stepanov, Su, Sun, Sun, Sun, Tam, Tang, Tang, Tian, Wang,
  Wang, Wang, Wang, Wang, Wang, Wang, Wang, Wang, Wang, Wang, Wang, Wang, Wang, Wang, Wang, Wang, Wang, Wang, Wang, Wang, Wei, Wei, Wei, Wen, Wu, Wu, Wu, Wu, Wu, Xi, Xia, Xia, Xiang, Xiao, Xiao, Xin, Xin, Xing, Xiong, Xu, Xu, Xu, Xu, Xue, Yan, Yan, Yan, Yang, Yang, Yang, Yang, Yang, Yang, Yang, Yang, Yang, Yao, Yao, Ye, Yin, Yin, You, You, Yu, Yuan, Yue, Zeng, Zeng, Zeng, Zha, Zhang, Zhang, Zhang, Zhang, Zhang, Zhang, Zhang, Zhang, Zhang, Zhang, Zhang, Zhang, Zhang, Zhang, Zhang, Zhang, Zhang, Zhang, Zhao, Zhao, Zhao, Zhao, Zhao, Zheng, Zhou, Zhou, Zhou, Zhou, Zhou, Zhou, Zhou, Zhu, Zhu, Zhu, Zhu, Zuo, \& Collaboration)}]{lhaaso_catalog}
---. 2024, The Astrophysical Journal Supplement Series, 271, 25, \dodoi{10.3847/1538-4365/acfd29}

\bibitem[{Dekker {et~al.}(2024)Dekker, Holst, Hooper, Leone, Simon, \& Xiao}]{tevhalos_let}
Dekker, A., Holst, I., Hooper, D., {et~al.} 2024, Phys. Rev. D, 109, 083026, \dodoi{10.1103/PhysRevD.109.083026}

\bibitem[{Fang \& Murase(2023)}]{kefeng}
Fang, K., \& Murase, K. 2023, The Astrophysical Journal Letters, 957, L6, \dodoi{10.3847/2041-8213/ad012f}

\bibitem[{{Giacinti} {et~al.}(2020){Giacinti}, {Mitchell}, {L{\'o}pez-Coto}, {Joshi}, {Parsons}, \& {Hinton}}]{gg}
{Giacinti}, G., {Mitchell}, A.~M.~W., {L{\'o}pez-Coto}, R., {et~al.} 2020, \aap, 636, A113, \dodoi{10.1051/0004-6361/201936505}

\bibitem[{Kaci \& Giacinti(2024)}]{me}
Kaci, S., \& Giacinti, G. 2024, Imprints of PeV cosmic-ray sources on the diffuse gamma-ray emission.
\newblock \doarXiv{2406.11015}

\bibitem[{Lipari \& Vernetto(2018)}]{lipari}
Lipari, P., \& Vernetto, S. 2018, Phys. Rev. D, 98, 043003, \dodoi{10.1103/PhysRevD.98.043003}

\bibitem[{Ma {et~al.}(2022)Ma, Bi, Cao, Chen, Chen, Cheng, Gong, Gu, He, Hou, Huang, Huang, Liu, Shchegolev, Sheng, Stenkin, Wu, Wu, Wu, Xiao, Yao, Zhang, Zhang, \& Zuo}]{angular_resolution}
Ma, X.-H., Bi, Y.-J., Cao, Z., {et~al.} 2022, Chinese Physics C, 46, 030001, \dodoi{10.1088/1674-1137/ac3fa6}

\bibitem[{Manchester {et~al.}(2005)Manchester, Hobbs, Teoh, \& Hobbs}]{atnf_1}
Manchester, R.~N., Hobbs, G.~B., Teoh, A., \& Hobbs, M. 2005, The Astronomical Journal, 129, 1993, \dodoi{10.1086/428488}

\bibitem[{{Martin} {et~al.}(2022){Martin}, {Tibaldo}, {Marcowith}, \& {Abdollahi}}]{Martin2022}
{Martin}, P., {Tibaldo}, L., {Marcowith}, A., \& {Abdollahi}, S. 2022, \aap, 666, A7, \dodoi{10.1051/0004-6361/202244002}

\bibitem[{Smith {et~al.}(2023)Smith, Abdollahi, Ajello, Bailes, Baldini, Ballet, Baring, Bassa, Gonzalez, Bellazzini, Berretta, Bhattacharyya, Bissaldi, Bonino, Bottacini, Bregeon, Bruel, Burgay, Burnett, Cameron, Camilo, Caputo, Caraveo, Cavazzuti, Chiaro, Ciprini, Clark, Cognard, Corongiu, Orestano, Crnogorcevic, Cuoco, Cutini, D’Ammando, de~Angelis, DeCesar, Gaetano, de~Menezes, Deneva, de~Palma, Lalla, Dirirsa, Venere, Domínguez, Dumora, Fegan, Ferrara, Fiori, Fleischhack, Flynn, Franckowiak, Freire, Fukazawa, Fusco, Galanti, Gammaldi, Gargano, Gasparrini, Giacchino, Giglietto, Giordano, Giroletti, Green, Grenier, Guillemot, Guiriec, Gustafsson, Harding, Hays, Hewitt, Horan, Hou, Jankowski, Johnson, Johnson, Johnston, Kataoka, Keith, Kerr, Kramer, Kuss, Latronico, Lee, Li, Li, Limyansky, Longo, Loparco, Lorusso, Lovellette, Lower, Lubrano, Lyne, Maan, Maldera, Manchester, Manfreda, Marelli, Martí-Devesa, Mazziotta, McEnery, Mereu, Michelson, Mickaliger, Mitthumsiri, Mizuno, Moiseev, Monzani, Morselli,
  Negro, Nemmen, Nieder, Nuss, Omodei, Orienti, Orlando, Ormes, Palatiello, Paneque, Panzarini, Parthasarathy, Persic, Pesce-Rollins, Pillera, Poon, Porter, Possenti, Principe, Rainò, Rando, Ransom, Ray, Razzano, Razzaque, Reimer, Reimer, Renault-Tinacci, Romani, Sánchez-Conde, Parkinson, Scotton, Serini, Sgrò, Shannon, Sharma, Shen, Siskind, Spandre, Spinelli, Stappers, Stephens, Suson, Tabassum, Tajima, Tak, Theureau, Thompson, Tibolla, Torres, Valverde, Venter, Wadiasingh, Wang, Wang, Wang, Weltevrede, Wood, Yan, Zaharijas, Zhang, \& Zhu}]{fermi_catalog}
Smith, D.~A., Abdollahi, S., Ajello, M., {et~al.} 2023, The Astrophysical Journal, 958, 191, \dodoi{10.3847/1538-4357/acee67}

\bibitem[{Sommers(2001)}]{attenuation}
Sommers, P. 2001, Astroparticle Physics, 14, 271, \dodoi{https://doi.org/10.1016/S0927-6505(00)00130-4}

\bibitem[{Vecchiotti {et~al.}(2022)Vecchiotti, Pagliaroli, \& Villante}]{vittoria}
Vecchiotti, V., Pagliaroli, G., \& Villante, F.~L. 2022, Commun Phys, 5, 161, \dodoi{10.1038/s42005-022-00939-7}

\bibitem[{Yan {et~al.}(2024)Yan, Liu, Zhang, Li, Yuan, \& Wang}]{neutrinos_tevhalo}
Yan, K., Liu, R.-Y., Zhang, R., {et~al.} 2024, Nat. Astron., 8, 628–636, \dodoi{https://doi.org/10.1038/s41550-024-02221-y}

\end{thebibliography}
\bibliographystyle{aasjournal}



\end{document}